\documentclass[12pt]{iopart}
\usepackage{graphicx}
\usepackage{bm}
\usepackage{bbm}
\usepackage{amsfonts}

\begin{document}

\title[Current fluctuations in a stochastic system of classical particles]{Current fluctuations in a stochastic system of classical particles with next-nearest-neighbor interactions}

\author{Sara Kaviani$^1$ and Farhad H. Jafarpour$^{1,2}$}
\address{$^1$ Bu-Ali Sina University, Physics Department, 65174-4161 Hamedan, Iran}
\address{$^2$ Laboratoire J.A. Dieudonn\'e, UMR CNRS 7351, Universit\'e de Nice Sophia-Antipolis, Parc Valrose, 06108 Nice Cedex 02, France} 
\ead{farhad@unice.fr}

\begin{abstract}
A totally asymmetric exclusion process consisting of classical particles with next-nearest-neighbor interactions has been considered 
on a 1D discrete lattice with a ring geometry. Using large deviation techniques, we have investigated fluctuations of particle current in two 
and three-particle sectors. In two-particle sector, we have obtained exact large deviation function for the probability distribution 
of particle current. In this sector, we have also calculated the the propensity of each configuration to exhibit atypical particle current. 
Analytical results for both scaled cumulant generating function of particle current and average particle current 
are compared with those obtained from cloning Monte Carlo simulations. Numerical results in three-particle sector have also been presented.
\end{abstract}
\section{Introduction}
In mathematical modeling of traffic flow one often considers a system of particles which move along a one-dimensional discrete lattice under 
certain dynamical rules~\cite{SCN}. If the average speed of all cars are assumed to be the same, the simplest model to consider is the 
asymmetric simple exclusion process (ASEP)~\cite{L,DEHP}; however, a more detailed modeling is required to investigate the effects of 
different speeds which is more realistic in the traffic flows~\cite{SZ,CDSS,SSN,RSSS,N}. 

One of the most related models to the traffic flow, which was investigated by Katz et al.~\cite{KLS} in 1984, is a stochastic lattice gas model 
under the influence of a uniform external biasing force. The model is defined as follows 
$$
\begin{array}{l}
0100 \longrightarrow 0010 \;\;\; \mbox{with rate}\;\;\;1+\delta\\
1100 \longrightarrow 1010 \;\;\; \mbox{with rate}\;\;\; 1+\epsilon \\ 
0101  \longrightarrow 0011 \;\;\; \mbox{with rate}\;\;\;1-\epsilon\\ 
1101  \longrightarrow 1011 \;\;\; \mbox{with rate}\;\;\;1-\delta 
\end{array}
$$
in which $1$ ($0$) stands for an occupied (vacant) lattice site. This model will be simplified to a toy model for the traffic flow if one considers $\delta=\epsilon$. 
The dynamical rules are now given by
\begin{equation}
\label{rules}
\begin{array}{l}
100 \longrightarrow 010 \;\;\; \mbox{with rate}\;\;\; r \\
101 \longrightarrow 011 \;\;\; \mbox{with rate}\;\;\; q  \; .
\end{array}
\end{equation}
This simplified model can show the slowing down of a car when encountering another car in a short distance for $q < r$. The model has been 
studied in~\cite{AS} where the authors have found the stationary probability distribution for both open and periodic boundary conditions. 
The stationary current-density relation of this model under periodic boundary condition has also been obtained in~\cite{AS,DS}. 
In the repulsive case $q<r$ the current-density relation becomes asymmetric which is more realistic according to the traffic data. 
In the symmetric case $q=r$ the totally ASEP in continuous time is recovered and the current-density shows a symmetric relation. For $q>r$ a 
traffic interpretation can not be obtained and the model can be assumed to describe the attractive short-range interaction between hard-core 
particles that are driven by an external force. 

Recently, much attention has been focused on the study of fluctuations of dynamical observables in stochastic processes with conditioned 
dynamics~\cite{BTG,JS0,CT,PH,BBT}. These studies include, finding those space-time trajectories which generate a rare value of the 
dynamical observable over a long period of time and also the unconditional dynamics, sometimes called the effective dynamics, whose 
statistics reproduce the fluctuations of the original process conditioned on the occurrence of a rare event. A variety of studies on finding 
such unconditional effective dynamics which turns an atypical value of a given dynamical observable into a typical value in the stationary 
state, have been performed ranging from driven-diffusive models of classical particles to the stochastic models of gene expression~\cite{E,JS5,V,HK}. 
For instance in~\cite{PSS}, the authors have shown that in the ASEP on a ring, conditioned on carrying a large flux, the particle experiences 
an effective long-range potential which has a simple form similar to the effective potential between the eigenvalues of the circular unitary ensemble in 
random matrices. 
The incident of dynamical phase transitions in both classical and quantum systems has made the subject more appealing~\cite{BD,BDGJL,LTW,BDL,HG,PGH,HEPG,VGG,JTS,SA,ZM,BKL,KS,TPGP,PCGP,HT}. The mathematical tools of large deviation 
theory provide a suitable framework to investigate these fluctuations and dynamical phase transitions~\cite{T9,LAW}. 

In this paper, we consider the process defined by~(\ref{rules}) and investigate the fluctuations of the particle current. We would like to understand 
how the particle system organizes itself microscopically when it generates atypical currents. In other words, we are interested in recognizing those
configurations which have more natural tendency to create certain particle current later. Our motivation for studying 
this toy model is that, not much is known about the fluctuation of the particle current in the systems with next-nearest-neighbor interactions. 
We believe that this study might shed more light on the subject specially when continuous dynamical transitions take place. 
Most of our analytical results are done in a two-particle sector, where only 
two particles exist in the system. For a system with three particles, we have done numerical calculations. Our results in the two-particle sector show that, in the
large system-size limit, the model possesses three different dynamical phases and that at the boundaries of these phases it undergoes 
second-order dynamical phase transitions. While the model in the stationary-state does not show long-range correlation, the particles interact 
through a long-range effective potential when an atypical current is produced. We have found that in order to observe an atypical value of the particle current less than 
the typical value in the steady-state, the particles prefer to attach to each other regardless of the values of $q$ and $r$. However, when atypically large values of the 
particle current is observed two different scenarios take place depending on the values of $q$ and $r$: in some cases, which will be discussed later in the paper, 
the particles prefer to stay as far from each other as possible while in other cases they prefer to stay a single lattice site apart from each other. 
At $q=r$ this model becomes identical to the totally ASEP where the particles move only in one direction. Our results, in the limit of high flux of 
particles, converge to those obtained in~\cite{PSS} in this limit. 

The validity of analytical results are verified by performing cloning Monte Carlo simulation~\cite{PH,GK,LT,TL,GKLT}. We have obtained scaled 
cumulant generating function (SCGF) of the particle current, which is a Legendre-Fenchel transform of the large deviation function of the particle 
current, and also its first derivative i.e. the average particle current in the two-particle sector. The simulation results show a good 
agreement with the analytical results.

This paper is organized as follows: In section~\ref{II} we start with the stochastic generator of the process and then explain how the stochastic generator of the 
effective Markov process can be obtained by solving the largest eigenvalue problem of a tilted generator. In section~\ref{III} we diagonalize the tilted
generator in a two-particle sector. We will also discuss those cases where the exact analytical results can be obtained. The effective potential between the two 
particles in the limit of high particle current will be discussed in section~\ref{IV}. The results obtained from the cloning Monte Carlo algorithm are brought in 
section~\ref{V}. In section~\ref{VI} we will briefly discuss the three-particle sector. Concluding remarks are brought in section~\ref{VII}.
 
\section{\label{II} Effective stochastic generator of a Markov process}
In this paper we study the fluctuations of the particle current, as a dynamical observable, in 
the system defined by the dynamical rules in~(\ref{rules}). By applying external forces to the original process one can enhance the probability of observing an atypical 
particle current in the system. However, we are not interested in arbitrary external forces that would make rare fluctuations of our physical observable 
typical. Instead we are looking for those very specific external forces that retain the spatio-temporal patterns of the original unforced process that generates these rare 
fluctuations by its own intrinsic random dynamics. 

It has been shown that an unconditional effective stochastic generator with the above mentioned properties, can be constructed as follows: Let us start with   
the time evolution of the probability vector of our Markov process. This is given by a Master equation which can be written, using the quantum Hamiltonian 
formalism, as~\cite{DS} 
$$
\frac{d}{dt} \vert P(t) \rangle =  {\cal H} \vert P(t)  \rangle 
$$
in which the stochastic generator ${\cal H}$ for the process defined by~(\ref{rules}) on a lattice of length $L$ with periodic boundary condition is given by
$$
{\cal H}={\cal H}_{1}+ {\cal H}_{2}- {\cal H}_{0} \; .
$$
${\cal H}_{1}=r \sum_{k=1}^{L}{s_{k}^{-} s_{k+1}^{+}v_{k+2}}$ and $ {\cal H}_{2}=q \sum_{k=1}^{L}{s_{k}^{-} s_{k+1}^{+}n_{k+2}}$ 
correspond to the diffusion of particles and ${\cal H}_{0}=\sum_{k=1}^{L}(r (n_{k} v_{k+1} v_{k+2}) + q  (n_{k} v_{k+1} n_{k+2}))$ gives the diagonal part 
of the stochastic generator ${\cal H}$. Here $n_{k}$ is the particle number operator and that $v_k=1-n_k$. $s^{+}_k$ and $s^{-}_k$ correspond to 
creation and annihilation of particles respectively and are the $SU(2)$ spin-$1/2$ ladder operators acting on the lattice site $k$. These operators 
have the following properties
$$
\langle \mathbbm{1} | s_{k}^{+}=\langle \mathbbm{1} | v_{k} \quad \mbox{and} \quad \langle \mathbbm{1} | s_{k}^{-}=\langle \mathbbm{1} | n_{k}
$$
where $\langle \mathbbm{1} | $ is defined as $\langle \mathbbm{1} | \equiv (1 \;\; 1)$\;.

It turns out that the effective stochastic generator ${\cal H}_{\mbox{eff}}(s)$
which generates an atypical value of particle current in its steady-state is given by a Doob's h-transform~\cite{JS0}
\begin{equation}
\label{eff}
{\cal H}_{\mbox{eff}}(s)= U  {\cal H}(s) U^{-1} - \Lambda^{\ast}(s)
\end{equation}
in which $\Lambda^{\ast}(s)$ is the largest eigenvalue of an associated tilted generator ${\cal H}(s)$. $\Lambda^{\ast}(s)$ is also the scaled cumulant generating function
of the particle current. Since there is no backward jump, the increment of the particle current is $+1$ for every particle jump, hence the non-conservative tilted 
generator ${\cal H}(s)$ is given by
\begin{equation}
\label{hs}
{\cal H}(s)=e^{s} ( {\cal H}_{1}+{\cal H}_{2})- {\cal H}_{0} 
\end{equation} 
with $\langle \tilde\Lambda^{\ast}(s) \vert {\cal H}(s)=\Lambda^{\ast}(s)\langle \tilde\Lambda^{\ast}(s) \vert$ and 
${\cal H}(s) \vert \Lambda^{\ast}(s) \rangle=\Lambda^{\ast}(s)\vert \Lambda^{\ast}(s) \rangle$. $U$ in~(\ref{eff}) is a diagonal matrix and its 
diagonal matrix elements are equal to the elements of $\langle \tilde\Lambda^{\ast}(s) \vert$.

Assuming that the particle current distribution has asymptotically a large deviation form as $e^{-tI(j)}$ in the limit of large $t$, where $t$ is the observation time and 
$j$ is the time-averaged current, the rate function $I(j)$ is related to $\Lambda^{\ast} (s)$ through a Legendre-Fenchel transform~\cite{T9}
\begin{equation}
\label{LFT}
I(j)=\sup_{s \in \mathbb{R}} \{ sj-\Lambda^{\ast} (s)  \} \; .
\end{equation}
The parameter $s$ connects the average particle current $j$ to $\Lambda^{\ast}(s)$ through 
\begin{equation} 
j=\frac{d}{ds}\Lambda^{\ast}(s) \; .
\end{equation}
$j$ is a monotonically increasing function of $s$ and it coincides at the point $s = 0$ with the stationary current of the process. On the other 
hand, a positive (negative) value of $s$ corresponds to an atypical current enhanced (reduced) with respect to the typical stationary current.
  
In the following section we will investigate the particle current fluctuations in the process defined by~(\ref{rules}) in a two-particle sector where 
only two particles exist on the lattice. In this regard, we need to diagonalize its corresponding tilted generator~(\ref{hs}). 

\section{\label{III} Two-particle sector}
According to~(\ref{LFT}), in order to calculate the large deviation function for the distribution of the particle current we need to 
find the largest eigenvalue of the tilted generator defined by~(\ref{hs}). It turns out that the eigenvalue problem 
$\langle \tilde\Lambda(s) \vert {\cal H}(s)=\Lambda(s)\langle \tilde\Lambda(s) \vert$ can
be solved analytically in the two-particle sector which is actually a one body problem in the relative coordinate. Let us consider
$$
 [{\cal H}(s)]^T \vert \tilde{\Lambda} (s) \rangle = \Lambda(s) \vert \tilde{\Lambda} (s) \rangle
$$  
in which $T$ stands for ``Transpose", and write  
\begin{equation}
\label{EV}
\vert \tilde{\Lambda} (s) \rangle =\sum_{1 \le x_1 < x_2 \le L} \psi(x_1,x_2) \vert x_1,x_2 \rangle 
\end{equation}
in which $x_1$ and $x_2$ are the positions of the particles. By applying the transpose of~(\ref{hs}) on~(\ref{EV}) we find the relations between 
$\psi(x_1,x_2)$'s. However, as a one body problem in the relative coordinate, everything will depend on the distance of the particles $l\equiv \vert x_2-x_1\vert$ 
($l=1$ means that the particles are at two consecutive lattice sites and so on.). In the two-particle sector $L$ is assumed to be an even number. 
In terms of the distance between the particles the elements of the left eigenvector $X_l\equiv \psi(x_1,x_2)$ satisfy the following 
recursion relations
\begin{eqnarray}
\label{eq1}
&&X_2=A X_1\; ,\nonumber\\
&&X_3=B X_1 \; ,\nonumber\\
&&X_{l} =C X_{l-1} -X_{l-2}\; \mbox{for}\; L\ge 8 \;\mbox{and }\; l=4,...,\frac{L}{2} \; ,\\
&&X_{\frac{L}{2}-l}=X_{\frac{L}{2}+l} \;\mbox{for} \; l=1,...,\frac{L}{2}-1 \nonumber
\end{eqnarray}
in which 
\begin{eqnarray}
\label{eq2}
&&A=\frac{\Lambda(s)+r}{r e^{s}},\nonumber\\
&&B=-\frac{q}{r}+A \frac{q}{r e^{s}}+A^{2},\\
&&C=\frac{\Lambda(s)+2r}{r e^{s}}.\nonumber\;
\end{eqnarray}
Considering $X_1$ as a free parameter and assuming 
\begin{equation}
X_{l}=Dz^{l}+E z^{-l} \;\; \mbox{for} \;\; l=2,\cdots,\frac{L}{2}
\end{equation}
we can use the first and second relations of~(\ref{eq1}) to find
\begin{eqnarray*}
&&D=\frac{Bz-A}{z^2(z^2-1)}X_{1}\;, \\
&&E=\frac{(A-Bz^{-1})z^4}{z^2-1}X_{1} \;. 
\end{eqnarray*}
Using the last relation of~(\ref{eq1}) we find $E/D=z^{L}$; therefore,
\begin{equation}
\label{lev}
X_{l}=D(z^{l}+z^{L-l}) \;\; \mbox{for} \;\; l=2,\cdots,\frac{L}{2} \; .
\end{equation}
Substituting~(\ref{lev}) in the third relation of~(\ref{eq1}) results in $C=z+z^{-1}$ which is equivalent to
\begin{equation}
\label{Lambda}
\Lambda(s)=r e^s(z+z^{-1})-2r \; .
\end{equation}
Finally using $E/D=Z^L$ and~(\ref{Lambda}) we find 
\begin{eqnarray}
\label{zeq}
z^{L-4}&=&-\frac{Az-B}{A z^{-1}-B} \\ 
&=&-\frac{(z+z^{-1}-e^{-s})(e^{-s}(1-\frac{q}{r})-z^{-1})+\frac{q}{r}}{(z+z^{-1}-e^{-s})(e^{-s}(1-\frac{q}{r})-z)+\frac{q}{r}} \;.\nonumber
\end{eqnarray}
One should, in principle, solve~(\ref{zeq}) and find the $z^\ast$ which maximizes~(\ref{Lambda}). From there the left 
eigenvector associated with the largest eigenvalue can be obtained using~(\ref{lev}). We have only been able to solve~(\ref{zeq})
in the large-$L$ limit. It is easy to see that~(\ref{zeq}) is invariant under $z \to z^{-1}$; therefore, $z$ can be consider as either 
a real number between $0$ and $1$ or a phase factor. Assuming $0 < z < 1$ then the numerator of~(\ref{zeq}), which is a second-order equation in 
$z$, goes to zero in the large system-size limit for certain values of $s$, $q$ and $r$. For other values of $s$, $q$ and $r$ the solution of~(\ref{zeq}) 
is a phase factor because for these values we find $\vert z \vert > 1$.  
A more detailed analysis of~(\ref{zeq}) in large-$L$ limit shows that the largest eigenvalue of the tilted generator is given by three 
different expressions depending on $\alpha\equiv q/r$. The second order derivative of the largest eigenvalue with respect to $s$ 
changes discontinuously at the boundaries of these three regions. 

\begin{figure}
\centering
\includegraphics[scale=0.6]{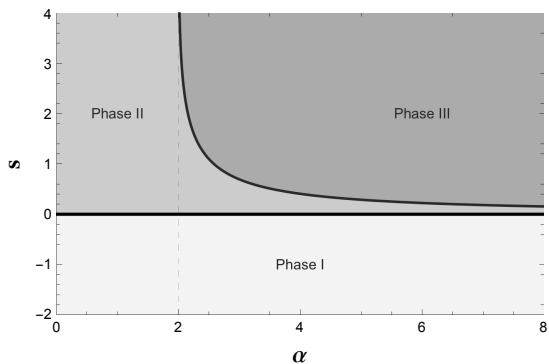}  
\caption{Phase diagram of the model in the $s-\alpha$ plane. The lines of second-order dynamical phase transitions are 
denoted as thick lines $s=0$ and $s=\ln ((\alpha-1)/(\alpha-2))$. }
\label{diagram}
\end{figure}

In FIG.~\ref{diagram} we have plotted the phase diagram 
of the model in $s-\alpha$ plane. Different phases are shown as areas with different shades of gray. The boundaries of different
phases are determined by two lines $s=0$ and $s=\ln((\alpha-1)/(\alpha-2))$ in $s-\alpha$ plane. 
In what follows we will discuss these phases by considering two different cases.

\subsubsection{The case $0 \le \alpha < 2$}
We have found that in the large system-size limit the largest eigenvalue is given by
\begin{equation}
\label{ev12}
\Lambda^{\ast}(s)=\left\lbrace
		\begin{array}{l}
			r(e^{2s}-1) \; \mbox{with} \; z^{\ast}=e^s \; \mbox{for} \; s \le 0 \;, \\ \\
			2r(e^s-1) \; \mbox{with} \; z^{\ast}=e^{\frac{i \pi}{L}} \; \mbox{for} \; s \ge 0 
		\end{array}
\right. \; .
\end{equation}

\begin{figure*}
\centering
\includegraphics[width=.45\linewidth]{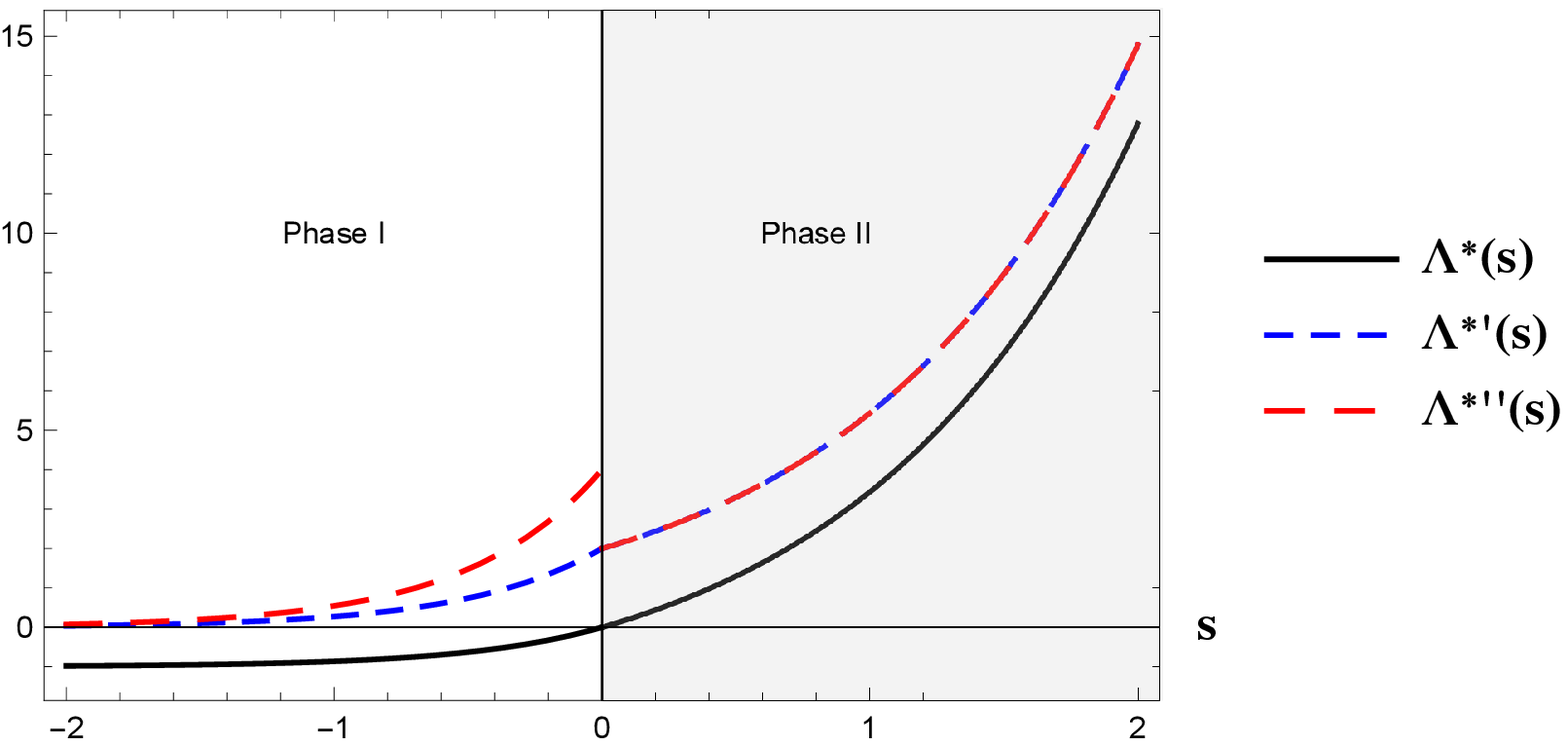}
\hfill
\includegraphics[width=.45\linewidth]{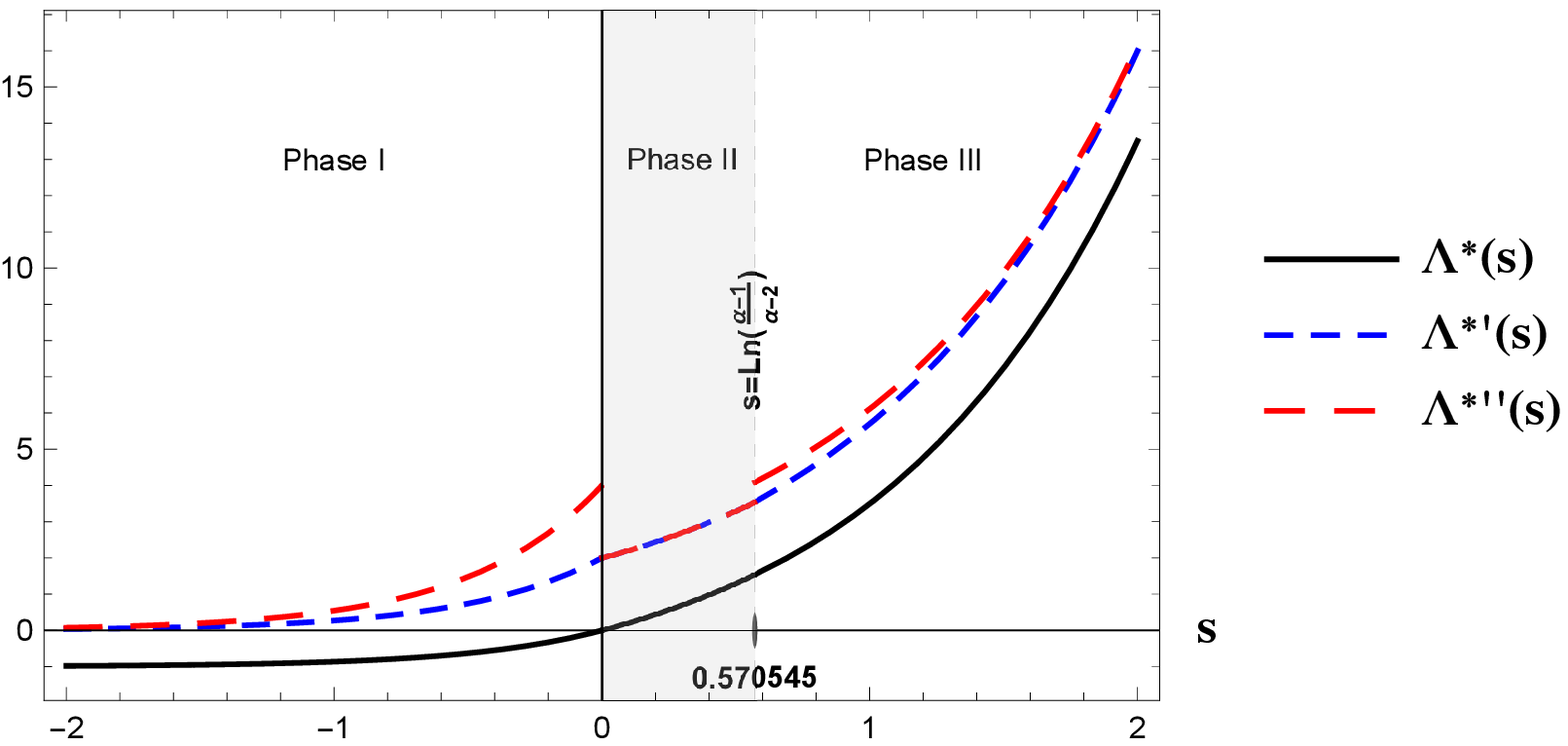}
\caption{Plot of~(\ref{ev12}) (left) and its first and second derivatives as a function of $s$ for $q=1.5$ and $r=1$. 
Note that for $s \ge 0$ both the first and the second derivatives lie on each other. Plot of~(\ref{ev123}) (right) and its first and second 
derivatives as a function of $s$ for $q=3.3$ and $r=1$. Note that for $0 \le s \le 0.570545$ both the first and 
the second derivatives lie on each other. In both figures the region where the first and second derivatives of $\Lambda^{\ast} (s)$ overlap is denoted in gray.
As can be seen the second derivative of $\Lambda^{\ast}(s)$ with respect to $s$ changes discontinuously in both cases indicating a second-order dynamical
phase transition.}
\label{lambda}
\end{figure*}

In FIG.~\ref{lambda} we have plotted $\Lambda^{\ast}(s)$ and its first and second derivatives as a function of $s$. These regions are 
denoted by Phase I and Phase II. As can be seen while both the largest eigenvalue and its first derivative are continuous at $s=0$, its 
second derivative with respect to $s$ changes discontinuously at that point. This is a sign for a second-order dynamical phase transition at $s=0$. 

Using~(\ref{LFT}) one can easily calculate the large deviation function for the particle current in this case. Straightforward calculations give 
$$
I(j)=\left\lbrace
		\begin{array}{l}
			r-\frac{j}{2}+\frac{j}{2}\ln (\frac{j}{2r}) \; \mbox{for} \; 0 \le j \le 2r \;, \\ \\
			2r-j+j\ln (\frac{j}{2r}) \; \mbox{for} \; j \ge 2r 
		\end{array}
\right. \; .
$$

\subsubsection{The case $\alpha \ge 2$}
In this case we have found, in the large system-size limit, that the largest eigenvalue of the tilted generator~(\ref{hs}) has the following behavior 
\begin{equation}
\fl
\label{ev123}
\Lambda^{\ast}(s)=\left\lbrace
		\begin{array}{l}
			r(e^{2s}-1) \; \mbox{with} \; z^{\ast}=e^s \; \mbox{for} \; s \le 0 \;, \\ \\
			2r(e^s-1) \; \mbox{with} \; z^{\ast}=e^{\frac{i \pi}{L}} \; \mbox{for} \; 0 \le s \le \ln({\frac{\alpha-1}{\alpha-2}}) \;,  \\ \\
			\frac{r}{2}(-2-\alpha+\alpha\sqrt{\frac{\alpha-1+4e^{2s}}{\alpha-1}})  \; \mbox{with} \; z^{\ast}=\frac{e^{-s}}{2}(1+\sqrt{\frac{\alpha-1+4e^{2s}}{\alpha-1}} )
			\; \mbox{for} \; s \ge \ln({\frac{\alpha-1}{\alpha-2}}) 
		\end{array}
\right. \; .
\end{equation}
In FIG.~\ref{lambda} we have plotted $\Lambda^{\ast}(s)$ and its first and second derivative with respect to $s$ in this case. The above regions are 
denoted by Phase I, Phase II and Phase III respectively. As in the previous case, one can see that while both the largest eigenvalue and its first derivative 
change continuously at $s=0$ and $s=\ln ((\alpha-1)/(\alpha-2))$, its second derivative with respect to $s$ changes discontinuously at these points. 

Similar to the previous case, the large deviation function for the particle current can be calculated exactly and the result is given by
$$
I(j)=\left\lbrace
		\begin{array}{l}
			r-\frac{j}{2}+\frac{j}{2}\ln (\frac{j}{2r}) \; \mbox{for} \; 0 \le j \le 2r \;, \\ \\
			2r-j+j\ln (\frac{j}{2r}) \; \mbox{for} \; 2r  \le j \le 2r ({\frac{\alpha-1}{\alpha-2}})  \;, \\ \\
			\frac{1}{2}(2+\alpha-j-\sqrt{\alpha^2+j^2}+ 
			      j\ln\frac{j(\alpha-1)(j+ \sqrt{\alpha^2+j^2})}{2\alpha^2})\; \mbox{for} \; j \ge 2r({\frac{\alpha-1}{\alpha-2}}) 
		\end{array}
\right. \; .
$$

The effective stochastic generator~(\ref{eff}), whose matrix elements are the effective rates associated with a given $s$, can be calculated using 
the largest eigenvalue of the tilted generator besides its corresponding left eigenvector in each dynamical phase. 

Following~\cite{JS0,EV} we define the effective potential between the particles as $U(l)=-2\ln X_l$ which can be calculated using~(\ref{lev}) and 
$z^\ast$ in each dynamical phase. $U(l)$ is actually a measure of ``the propensity of a configuration, in which the particles 
are $l$ lattice sites apart, to generate certain particle current in future''~\cite{JS0}.

\begin{figure}
\centering
\includegraphics[scale=0.6]{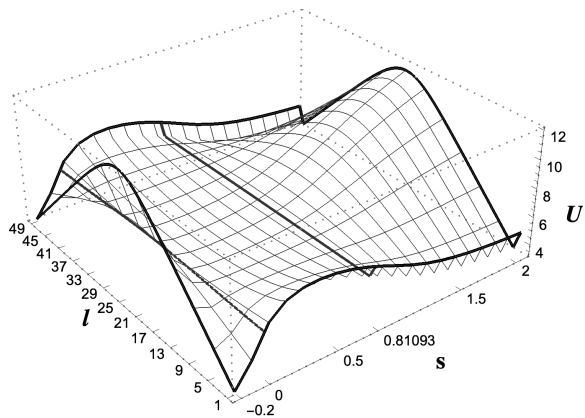}  
\caption{3D plot of $U(l)$ as a function of $l$ and $s$ at $\alpha=2.4$ and $L=50$. It can bee seen that the locus of the minimum of 
$U$ changes from $l=1$ (in Phase I) to $l=L/2$ (in Phase II) to $l=2$ (in Phase III) as $s$ is increased. At $s=0$ the effective potential is flat. 
At $s=\ln((\alpha-1)/(\alpha-2))=0.81093$ we see a steeply rising effective potential. At $s=0$ the effective potential is a straight line. For more detail see inside the text. }
\label{3D}
\end{figure}

In FIG.~\ref{3D} we have plotted numerically obtained $U(l)$ as a function 
of $l$ and $s$ at $\alpha=2.4$ for a system of size $L=50$. Our numerical analyses show that in Phase I ($s\le 0$) the locus of minimum of the effective 
potential is at $l=1$ (and equivalently $l=L-1$) where the particles are on two consecutive lattice sites. This means that the ``propensity" of this configuration to 
generate lower-than-typical particle current in future is higher than the other configurations. Using $z^\ast=e^s$ in the large system-size limit, we 
find $U(l)\cong 2 \vert s \vert l$ for $1 \le l \le L/2$ in this phase. At the boundary of Phase I and Phase II, i.e. on the line $s=0$, the effective potential is completely flat. 
This comes from the fact that at  $s=0$ the left eigenvector of the tilted generator is given by the summation vector $(1\;1\;1\; \cdots\; 1\; 1)$ 
and hence, all configurations have equal chance to produce the typical current in future. 
In Phase II the locus of the minimum of $U(l)$ it is at $l=L/2$ where the particles are at maximum distance from each other. In Phase III the locus 
of the minimum of $U(l)$ is at $l=2$ (and equivalently $l=L-2$) where the particles are a single lattice site apart from each other. 

\section{\label{IV} Large-$s$ limit}
Let us consider the limit of very large $s$, corresponding to high particle current. 
In this limit~(\ref{zeq}) becomes
\begin{equation}
\label{eq3}
z^{L-4}=-\frac{1+z^{-2}-\alpha}{1+z^{2}-\alpha} \; .
\end{equation}
  
\begin{figure}
\centering
\includegraphics[scale=0.5]{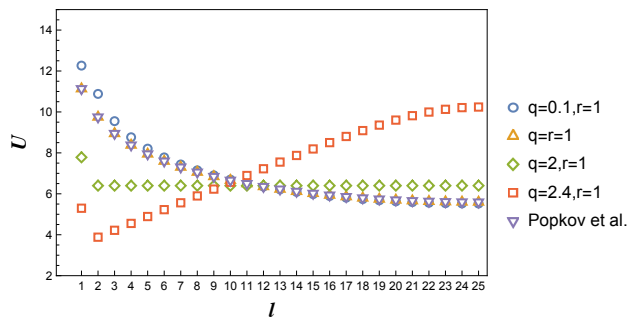}  
\caption{Plot of $U(l)$ versus $l$ for a system of length $L=50$ and various values of 
$q$ and $r$ at $s=100$. For $q/r<2$ the minimum of the effective potential is at $l=L/2$ while for $q/r>2$ it is at $l=2$. At $q=r$ our 
results coincide with those obtained in~\cite{PSS}. Note that at $q/r=2$, $U(l)$ steeply rises at the boundaries.}
 \label{potential}
\end{figure}

In FIG.~\ref{potential} we have plotted the effective potential between the particles as a function of $l$ ($1\le l \le L/2$) for different 
values of $\alpha$ at a large $s$. Using $X_{L/2-l}=X_{L/2+l} \;\mbox{for} \; l=1,...,L/2-1$ one can calculate $U(l)$ for $l=L/2+1,L-1$. Strong dependency 
of $X_l$'s to $l$, which can be seen in FIG.~\ref{potential}, means that even if the dynamical rules in~(\ref{rules}) are local, the matrix elements 
of~(\ref{eff}) depend on $l$, which implies that the effective dynamics is long-range.  

For $0 \le \alpha < 2$ (Phase II), $U(l)$ has a single minimum; however, as $\alpha$ increases this minimum
broadens so that at $\alpha=2$ we see a steeply rising function at the boundaries. At $\alpha=2$ the probability of creating high particle current is low when the particles 
are at two consecutive lattice sites, while other configurations have equal probability to produce high particle current. At $\alpha=1$ we find from~(\ref{eq3}) that
$z^{L}=-1$ and the largest eigenvalue is given by
\begin{equation}
z^{\ast}=e^{\frac{i \pi}{L}} \; .
\end{equation}
The elements of the left eigenvector~(\ref{lev}) become
\begin{equation}
X_{l}\propto \sin (\frac{l \pi}{L}) \;\; \mbox{for} \;\;l=1,\cdots , \frac{L}{2} \;.
\end{equation}
This result coincides with the one that has already been obtained in~\cite{PSS}. This is not surprising since the dynamical rules of this model and 
those of the totally ASEP  ($p=1$ and $q=0$ in~\cite{PSS}) become identical at this point. At $\alpha=1$ the particles should be at the 
farthest possible distance from each other in order to produce high particle current in the system. 

For $\alpha > 2$ (Phase III) the effective potential has two minima at $l=2$ and $l=L-2$ corresponding to the configurations when the particles are one site 
apart from each other. These configurations generate high particle current while the probability of generating high particle current by other particle 
configurations is low. 

\section{\label{V} Cloning Monte Carlo}
\begin{figure}
\includegraphics[width=.45\textwidth]{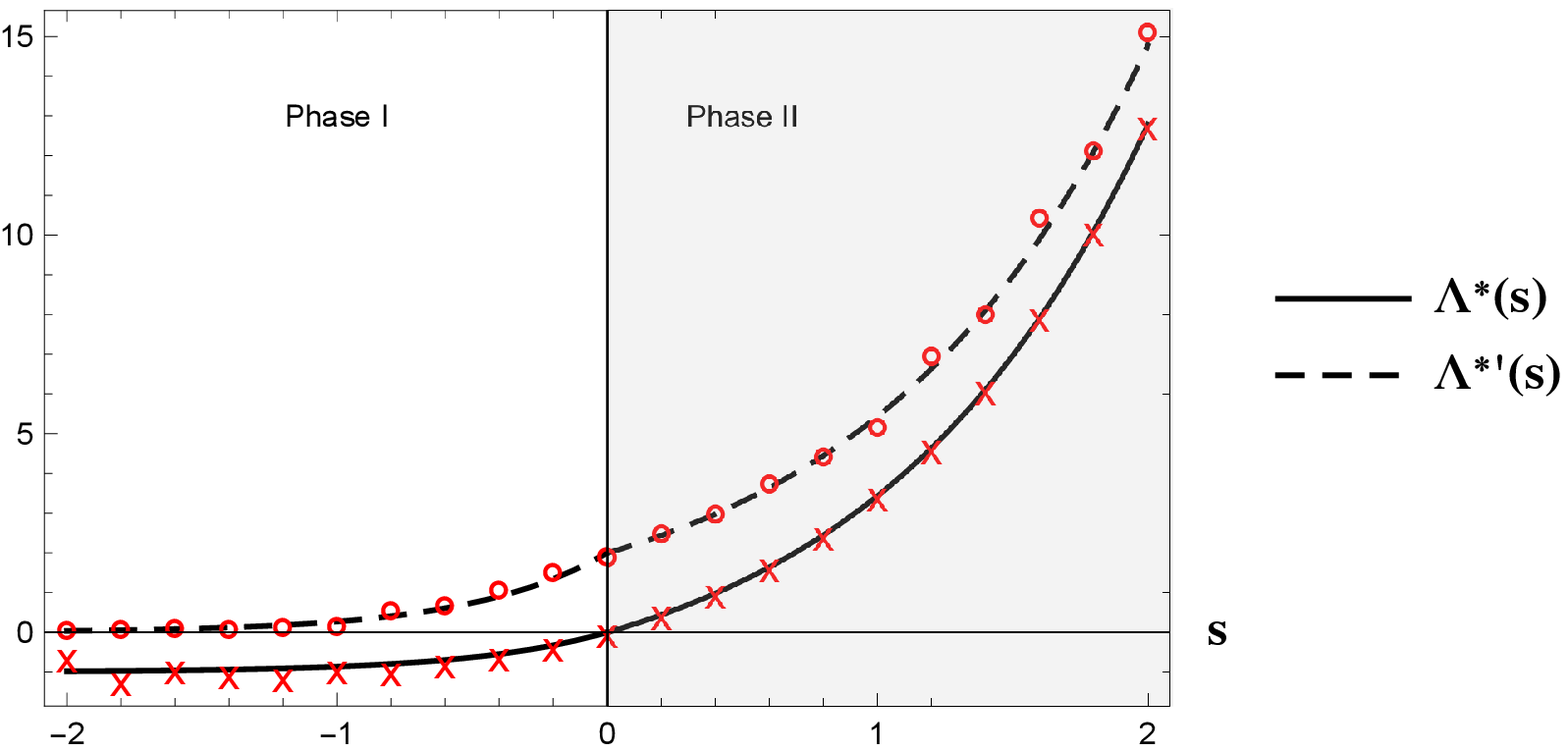}
\hfill
\includegraphics[width=.45\textwidth]{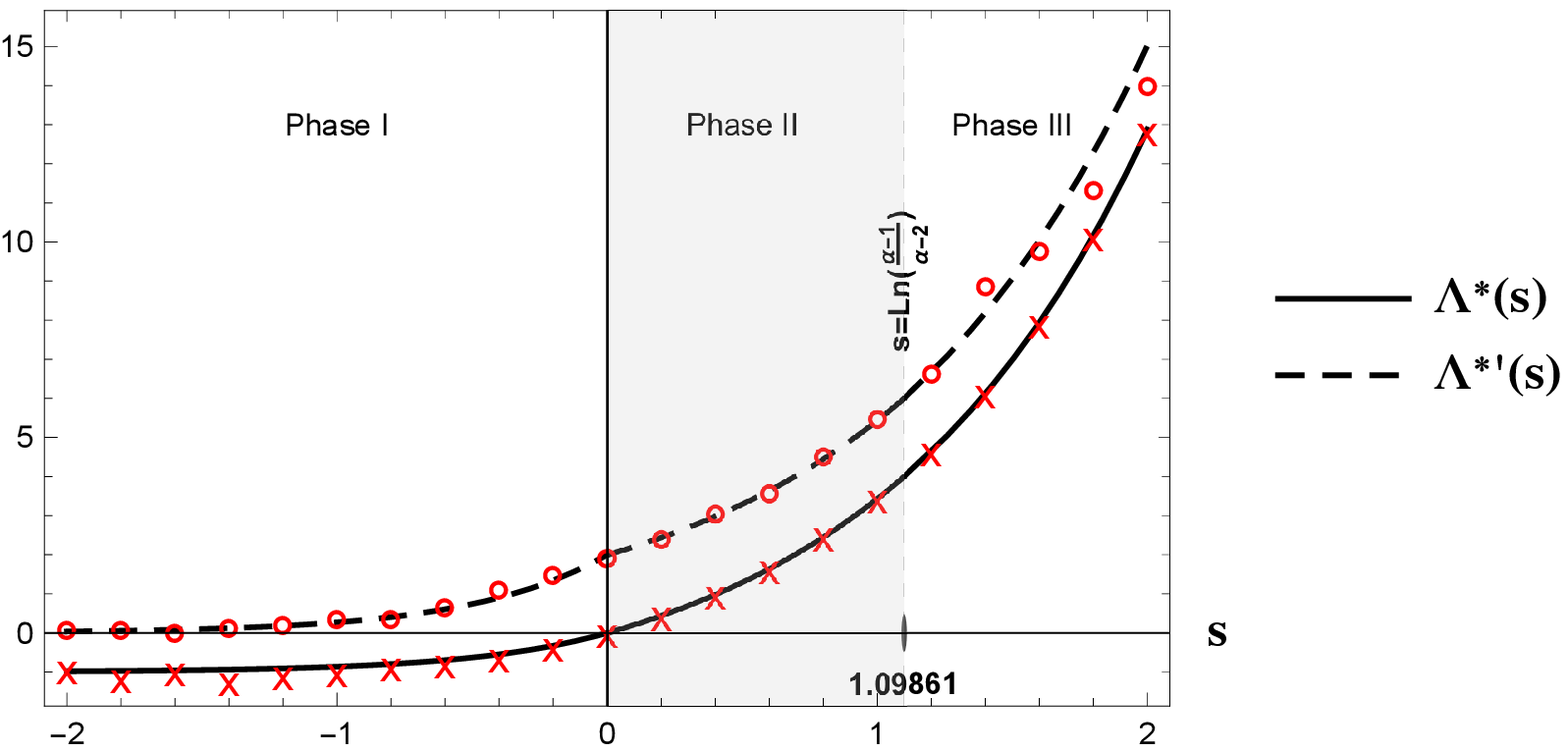}
\caption{Solid and dashed lines are plots of $\Lambda^{\ast}(s)$ and $\Lambda^{\ast\prime}(s)$ as a function of $s$ 
(obtained from~(\ref{ev12}) and~(\ref{ev123})) for $q=1.5r=1.5$ (left) and $q=2.5r=2.5$ (right) respectively.
Symbols are the results of cloning Monte Carlo for the same values of the parameters in each figure and $L=200$.}   
\label{EVS}
\end{figure}

In order to verify our analytical findings for the \textit{dynamical free energy} $\Lambda^{\ast}(s)$ and its first derivative, we have
performed the cloning Monte Carlo simulation. Using the population dynamics method in continuous time 
explained in~\cite{TL}, we have calculated $\Lambda^{\ast}(s)$ as a function of $s$. We start with $10^3$ clones of the original 
system of length $L=200$. The results are compared with~(\ref{ev12}) and~(\ref{ev123}). As can be seen in FIG.~\ref{EVS} the 
agreement between analytical result and simulation is significant. We have also used the same technique 
to obtain the average particle current $j=d\Lambda^{\ast}(s)/ds$. The averages are taken over the whole dynamical trajectories 
in the long time limit which is less noisy~\cite{TL}. The results for the average particle current, or equivalently 
the first derivative of the dynamical free energy with respect to $s$, are also brought in FIG.~\ref{EVS}. 

\section{\label{VI}Three-particle sector}
In this sector the effective potential is calculated numerically in the large-$s$ limit. Let us denote the position of the particles on the lattice by $x_i$ with $i=1,2,3$.
Because of the translational symmetry we only need two labels to fully determine the positions of the particles. These labels are called $l_1=|x_2-x_1|$ and 
$l_2=|x_3-x_2|$. 

\begin{figure*}
\centering
\includegraphics[width=0.3\textwidth]{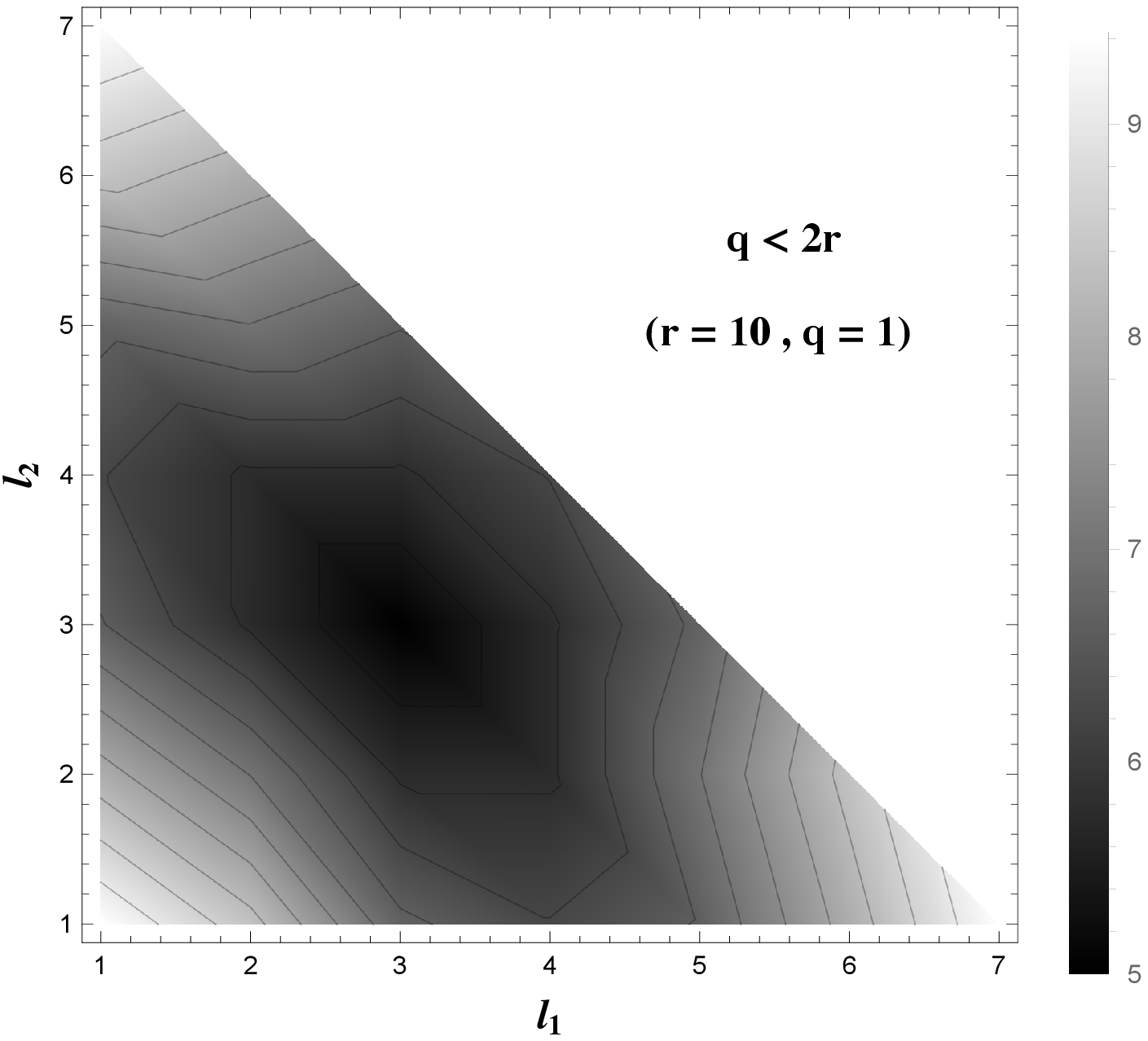}
\hfill
\includegraphics[width=0.3\textwidth]{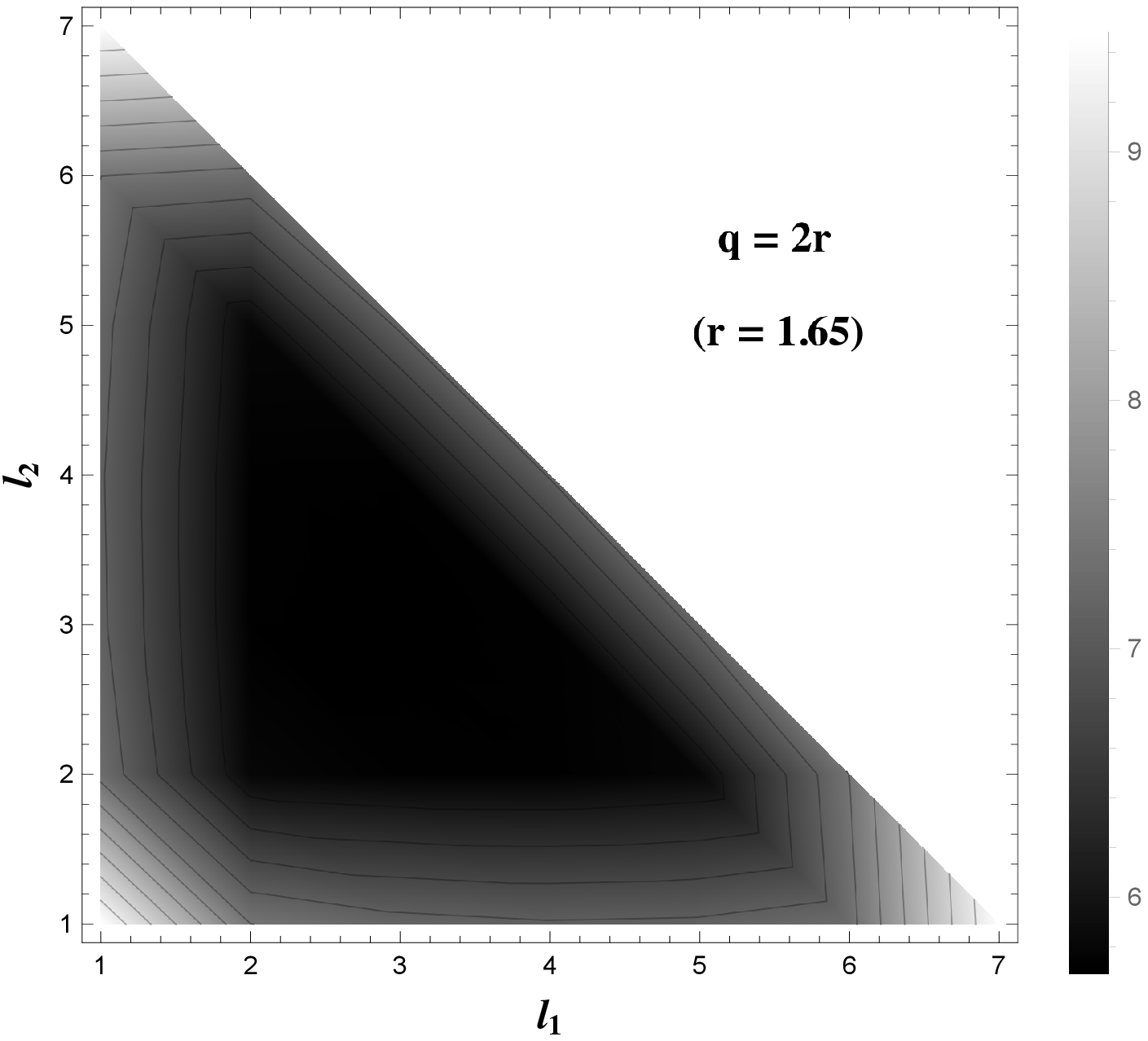}
\hfill
\includegraphics[width=0.3\textwidth]{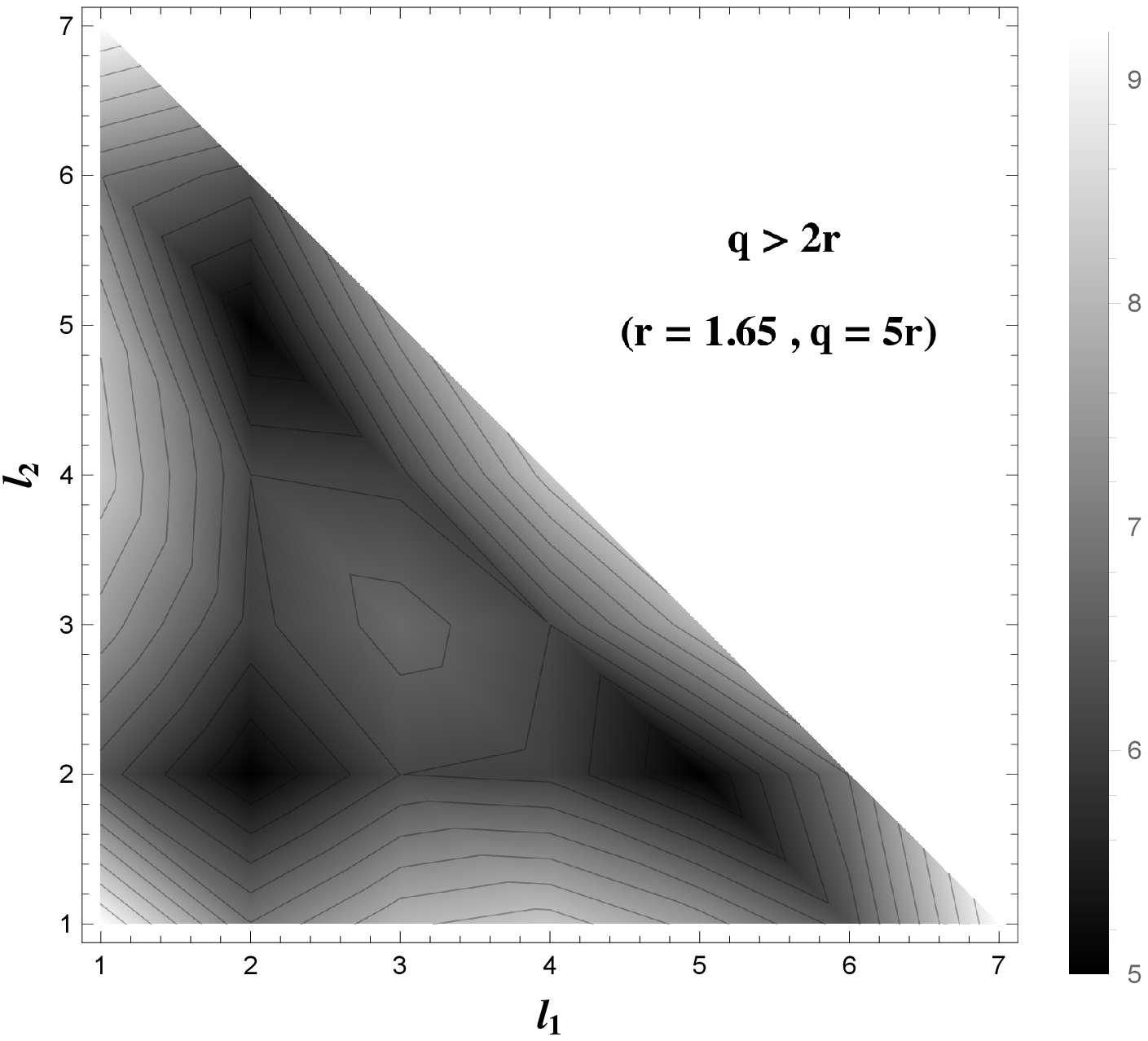}
\caption{Density plot of $U(l_1,l_2)$ for a system of length $L=9$, $s=500$, $\alpha=0.1$, $\alpha=2$ and $\alpha=5$
from left to right respectively. In order to generate high particle flux, for $\alpha > 2$ the particles prefer to stay a single lattice site apart while for $\alpha < 2$ they 
prefer to stay as far from each other as possible.}
 \label{3particle}
\end{figure*}

For $\alpha<2$, for instance $\alpha=1$ when the dynamical rules turn the model to the totally ASEP with
$p=1$ and $q=0$ in~\cite{PSS}, the most probable configuration which generates high particle current is the one in which the particles try to
stay far from each other (maximum distance). In a system of length $L=9$ high particle current is highly probable to be generated by $l_1=l_2=3$. 
This can be seen in FIG.~\ref{3particle} in which $U(l_1,l_2)$ is plotted as a function of $l_1$ and $l_2$ for $\alpha=0.1$.

At $\alpha=2$, similar to the two-particle sector, the effective potential $U(l_1,l_2)$ has a broad minimum. This can also be seen in FIG.~\ref{3particle}
where the effective potential has a broad minimum and high particle current can be generated equally likely by any configurations of particles 
except those in which the particles are at two consecutive lattice sites ($l_1=1$, $l_2=1$), ($l_1=7$, $l_2=1$) or ($l_1=1$, $l_2=7$).

Finally, as $\alpha$ increases beyond $\alpha=2$ the high particle current is generated by those configurations in which the particles form a cluster and
are a single site apart from each other. This can be seen in FIG.~\ref{3particle} for $\alpha=5$ where the most probable configurations which 
generate high particle current are ($l_1=2$, $l_2=2$), ($l_1=5$, $l_2=2$) or ($l_1=2$, $l_2=5$). 

\section{\label{VII} Concluding remarks}
In this paper we have studied the fluctuations of the particle current in a driven-diffusive system of classical particles 
with three-site interactions. We have been able to calculate the large deviation function for the probability distribution
of particle current besides the effective potential between the particles in the two-particle sector. We have found that the 
phase diagram of the process has three different phases. The largest eigenvalue of the tilted generator or 
the dynamical free energy of the system behaves differently in each dynamical phase. Moreover, the second
derivative of the dynamical free energy changes discontinuously at the boundaries of different phases. Calculating the  
effective potential for a given configuration allows us to find the propensity or natural tendency of that configuration to exhibit flux in the
future. We have been able to find those configurations which generate high particle current in each phase. In the three-particle
sector our results are numerical. We have calculated the effective potential by numerical diagonalization of the tilted generator
of the process. It turns out that, in three-particle sector, the way that the particles organize themselves to generate high current is similar   
to that in the two-particle sector in the sense that the particles should be either a single site apart or completely far from each other depending on $\alpha$.  
We have found that the same is true as long as the density of the particles is less than one-half. However, it would be of great interest to see 
how the system behaves, when it generates rare values of the particle current, as the number of particles increases. 
An approach to solve this open problem can be application of the macroscopic fluctuation theory.

\section*{Acknowledgments}
FHJ would like to thank the hospitality of the Laboratoire J.A. Dieudonn\'e at Universit\'e de Nice where part of this work has been done. The 
authors would like to thank the anonymous referee for his/her enlightening comments and suggestions. 
 
\section*{References}


\end{document}